\documentclass[journal]{IEEEtran}
\usepackage[latin9]{inputenc}
\usepackage{latexsym,caption}
\usepackage{amsthm}
\usepackage{amssymb}
\usepackage{setspace}
\usepackage{amsmath}
\usepackage{multirow}
\usepackage{xcolor}

\usepackage[pdftex]{graphicx}
\usepackage[colorlinks = true, linkcolor = green, urlcolor  = black, citecolor = blue, anchorcolor = blue]{hyperref}

\ifCLASSINFOpdf
\else
\fi
\begin{document}
%

% \title{4D Sleipner Data Interpolation and Extrapolation}

\title{Connect the Dots: In Situ 4D Seismic Monitoring of CO$_2$ Storage with Spatio-temporal CNNs}

% author names and IEEE memberships
% note positions of commas and nonbreaking spaces ( ~ ) LaTeX will not break
% a structure at a ~ so this keeps an author's name from being broken across
% two lines.
% use \thanks{} to gain access to the first footnote area
% a separate \thanks must be used for each paragraph as LaTeX2e's \thanks
% was not built to handle multiple paragraphs
%
\author{Shihang Feng$^\mathrm{2, *}$, Xitong Zhang$^\mathrm{1, 3}$, Brendt Wohlberg$^\mathrm{2}$, Neill Symons$^\mathrm{1}$,  and Youzuo Lin$^\mathrm{1, *}$
\thanks{\textbf{1}: Earth and Environmental Sciences Division, Los Alamos National Laboratory, Los Alamos, NM, 87544 USA.}% <-this % stops a space
\thanks{\textbf{2}: Theoretical Division, Los Alamos National Laboratory, Los Alamos, NM, 87544 USA.}
\thanks{\textbf{3}: Department of Computational Mathematics, Science and Engineering, Michigan State University, East Lansing, MI 48824 USA.}
\thanks{$^\mathrm{*}$Correspondence to:  S. Feng (shihang@lanl.gov) and Y. Lin (ylin@lanl.gov).}}

\markboth{IEEE Transactions on Geoscience and Remote Sensing}%
{Shell \MakeLowercase{\textit{et al.}}: Bare Demo of IEEEtran.cls for Journals}

% make the title area
\maketitle

% As a general rule, do not put math, special symbols, or citations
% in the abstract or keywords.
\begin{abstract}
4D seismic imaging has been widely used in CO$_2$ sequestration projects to monitor the fluid flow in the volumetric subsurface region that is not sampled by wells. Ideally, real-time monitoring and near-future forecasting would provide site operators with great insights to understand the dynamics of the subsurface reservoir and assess any potential risks. However, due to obstacles such as high deployment cost, availability of acquisition equipment, exclusion zones around surface structures, only very sparse seismic imaging data can be obtained during monitoring. That leads to an unavoidable and growing knowledge gap over time. The operator needs to understand the fluid flow throughout the project lifetime and the seismic data are only available at a limited number of times. This is insufficient for understanding the reservoir behavior. To overcome those challenges, we have developed spatio-temporal neural-network-based models that can produce high-fidelity interpolated or extrapolated images effectively and efficiently. Specifically, our models are built on an autoencoder, and incorporate the long short-term memory (LSTM) structure with a new loss function regularized by optical flow. We validate the performance of our models using real 4D post-stack seismic imaging data acquired at the Sleipner CO$_2$ sequestration field. We employ two different strategies in evaluating our models. Numerically, we compare our models with different baseline approaches using classic pixel-based metrics. We also conduct a blind survey and collect a total of 20 responses from domain experts to evaluate the quality of data generated by our models. Via both numerical and expert evaluation, we conclude that our models can produce high-quality 2D/3D seismic imaging data at a reasonable cost, offering the possibility of real-time monitoring or even near-future forecasting of the CO$_2$ storage reservoir. 

\end{abstract}

% Note that keywords are not normally used for peer-review papers.
\IEEEpeerreviewmaketitle

\section{Introduction}

\IEEEPARstart{T}{o} assess the safety and effectiveness of CO$_2$ sequestration projects, effective monitoring techniques are required to build a 4D~(3D in space and 1D in time) digital subsurface models that accurately matches the CO$_2$ injection over a significant period of time~(e.g. 10 - 200 years). Among all available geophysical monitoring techniques~(such as seismic, EM, gravity, ERT, etc.), 4D seismic imaging is one of the most effective and reliable techniques for monitoring and verification of CO$_2$ sequestration projects~\cite{Ma-2016-Geophysical}. However, due to insufficient data, in-situ 4D seismic monitoring of the reservoir changes is technically challenging or even infeasible.

The data scarcity is mainly caused by the practical limitations of high cost, availability of equipment, exclusion zones around surface structures, etc., that makes it difficult to deploy sufficient frequency of data acquisition~\cite{koster2000time}. For example, consider the world's first industrial-scale CO$_2$ storage operation project, Sleipner, a well-studied area in the North Sea~\cite{chadwick2004geological}. Since October 1996, Equinor, the site operator, has injected about 1 million tonnes~(Mt) of super-critical CO$_2$ per year~\cite{chadwick2010quantitative}. In order to monitor the migration of injected CO$_2$ and plume geometry development, comprehensive 3D-time lapse seismic surveys have been carried out. Due to practical limitations, data were acquired only in a limited number of years~(in 1994, 1999, 2001, 2002, 2004, 2006, 2008, 2010, 2013, and 2016). The missing data in between creates an unavoidable knowledge gap, which hinders the in-situ monitoring of the CO$_2$ migration. 
To improve the security and applicability of the CO$_2$ sequestration project, these missing data are required to be estimated for real-time monitoring, which is technically challenging.

%% Some in-between interpolation techniques used in CV

Recently, similar challenges and tasks have been actively studied in the computer vision community for generating new samples from a sequence of temporally-dependent images, namely, ``in-between image interpolation''~\cite{Generating-2020-Cristovao, Super-2018-Jiang, zuckerman2020across}. Broadly speaking, there are two categories of interpolation approaches, i.e., those based on pixels and those based on latent representations. Linear interpolation is a typical example of a pixel-based approach. Pixel-based methods usually cause problems when input images are temporally far apart, which would lead to a loss of temporal dependence and create errors like cross-dissolve effects~\cite{oring2020faithful}. On the other hand, by leveraging  semantic information, the latent representation-based approaches alleviate the problems of pixel-based models to some extent. However, directly applying those existing in-between image interpolation techniques to our 4D seismic imaging data is not trivial due to the differences in the nature of the problems as well as the spatio-temporal characteristics. 

%% Talk something about post-stack seismic imaging data and its spatio-tempoal characteristics. Also Say what would be the challenges

4D post-stack seismic imaging~(i.e., time-lapse 3D imaging) leverages repeating seismic surveys in succession. Via time-lapse 3D imaging techniques, time shifts and amplitude changes in the data can be measured and further interpreted as the changes in the reservoir and overburden~\cite{vasco2004seismic, macbeth2020methods, lumley2004business}. In a typical CO$_2$ sequestration project such as Sleipner, the seismic data can change rapidly as the CO$_2$ injection proceeds. This is due to the large perturbation of subsurface velocity as the fluid changes from pure brine to brine with low CO$_2$ saturation, while the velocity changes much less as CO$_2$ saturation increases further~\cite{furre201720}. 4D post-stack seismic imaging data can simultaneously detect  spatial and temporal changes~\cite{chadwick20054d} if the sampling is sufficiently dense. Spatially, the CO$_2$ plume in post-stack imaging with the Sleipner data is characterized as several bright sub-horizontal seismic reflections within the reservoir unit. The seismic reflections mostly come from thin layers of CO$_2$ trapped beneath very thin intra-reservoir mudstones and the reservoir caprock and the push-down is caused by the decrease of seismic velocity due to CO$_2$-saturated rock. Temporally, the migration and development of CO$_2$ plume over time is characterized by repeating surveys at relatively sparse times. Hence, filling the knowledge gap for post-stack seismic imaging would need to respect the spatio-temporal characteristics of the existing sparse measurements without violating the physics. In the problem in-situ monitoring, we need to generate in-between images (interpolation) and also to forecast new images from a few past measurements (extrapolation) which is more technically challenging. 

%% Now, say our methods. 

To overcome these aforementioned challenges, we propose a deep neural network-based interpolation/extrapolation model incorporating domain knowledge to ``connect the dots'' between the seismic data in years. Our method builds on an autoencoder to obtain latent variables of input sequential images. To further enhance the temporal correlation, we not only incorporate the long short-term memory (LSTM)~\cite{Hochreiter-1997-Long} but also use a new loss function regularized with optical flow~\cite{li2020complete}. Particularly, the LSTM structure is used to recognize the sequential and temporal dynamics, which are leveraged for forecasting. The optical flow describes the spatial shifting of objects in time-varying images. We use it as a regularization term to provide an additional layer of constraints when generating new images. 

We validate the performance of our techniques using both real 2D and 3D post-stack seismic imaging data acquired from the Sleipner project. Similar to previous work, we first compare our approaches to baseline methods by using standard per-pixel error metrics. However, classic per-pixel measures are insufficient for assessing structured data such as images~\cite{Zhang-2018-Unreasonable}. We carry out an expert evaluation test to further justify the quality of our generated images. Via both standard metrics and expert evaluation, we show that our model outperforms existing approaches and effectively and efficiently produces realistic images that could facilitate in-situ monitoring of CO$_2$ changes. 

This paper is organized into the following sections. After the introduction in the first section, we provide related work in data interpolation from both the seismic exploration and computer vision communities. We then present the foundation and theory including 3D time-lapse seismic imaging and Sleipner data set, autoencoder, and optical flow in the third section. Our new proposed interpolation and extrapolation networks are provided in the fourth section. The fifth section shows numerical results using the Sleipner data. A discussion of our interpolation and extrapolation networks is presented in the sixth section. Finally, the conclusion is given in the last section.

\section{Related Works}

\subsection{Interpolation and Extrapolation in Seismic Exploration}

Seismic interpolation is an important research topic in the field of seismic data processing. The prediction filters in the F-X domain~\cite{spitz1991seismic}, support vector regression (SVR)~\cite{jia2017can,lu2018auto}, U-net~\cite{mandelli2019interpolation} and CycleGAN~\cite{kaur2019seismic} had been applied to reconstruct seismic data from under-sampled or missing traces in the trace interpolation. Interpolation methods have been studied to obtain continuous information between 4D seismic surveys. Linear and Lagrange interpolation of the seismic response were used for calculating the onset time map~\cite{liu2020integration}. An amplitude interpolation method based on a probability formulation was developed to estimate the seismic data between the infrequent seismic survey~\cite{kiaer2016calendar}. Although these methods can produce smooth transitions between the repeated survey, interpolation uncertainty increases with the time gap between consecutive seismic surveys. 

Prediction of seismic data is also important in the study of 4D seismic survey. A comprehensive analysis of the geophysical data and geology is usually required to increase the confidence of the prediction~\cite{lindeberg2001prediction, hoversten2003pressure}. The amplitude interpolation method had been extended to predict the seismic data, but a large uncertainty exists with the extrapolation~\cite{kiaer2016calendar}. 
With some physical regularization such as optical flow, we can mitigate the uncertainty problem in our extrapolation network.
\begin{table*}[ht]
\centering
\caption{The total amount of CO$_2$ injected volume along with the time}
\label{table_aerial_slic4}
\begin{tabular}{cccccccccc}
\hline
&&\multicolumn{8}{c}{\textbf{Year}} \\ 
&\%& 1994&  1999    &   2001     &   2002     &     2004    &   2006 & 2008 & 2010 \\\cline{2-10}
&Mass (Mt) & 0      & 2.3 & 4.2 & 5.1  & 6.9 & 8.4 & 10.1 & 12.0  \\ 
 \hline
 \label{tab:inject}
\end{tabular}
\end{table*}
\subsection{Interpolation in Computer Vision}

Image interpolation has been the subject of intense study in computer vision. Existing methods can be generally categorized as pixel-based or latent-representation based~\cite{Generating-2020-Cristovao}. The former performs a simple manipulation of pixels by considering the motion of the objects from the pixel level. The latter learns a feature-based representation of object motion from the semantic level by convolving input images with spatially adaptive kernels. 

% Pixel-based interpolation methods have been extensively studied. ~\citet{iterative-1981-Lucas}

Pixel-based interpolation between images introduces ghosting artifacts, a sign of departure from the underlying manifold~\cite{upchurch2017deep}. Two directions had been studied to overcome the limitations of those approaches. The first direction is the motion-compensated frame methods~\cite{Motion-2010-Wang, low-2005-Zhai} and the second is the phase-based methods~\cite{Phase-2015-Meyer, Phase-2013-Wadwa}. The motion-compensated approaches are based on the assumption that motion in images is smooth and continuous. They estimate the motion based on previous and current images so that the in-between images can be generated by averaging the pixels in the images indicated by motion vectors~\cite{Motion-2010-Wang, low-2005-Zhai}. The phase-based methods assume that small motion can be encoded using phase shift~\cite{Phase-2015-Meyer, Phase-2013-Wadwa}.

With recent advances in neural networks, the latent-representation-based approaches had shown promising results in the task of in-between image interpolation~\cite{PhaseNet-2018-Meyer, berthelot2018understanding, sainburg2018generative, FlowNet-2017-Ilg, white2016sampling}. Due to the capability of disentangling important factors of variation in the dataset, those latent-representation-based approaches are particularly effective in producing semantically smooth interpolations~\cite{white2016sampling, sainburg2018generative}.  A critic was introduced in the training of the interpolation network to generate more realistic new samples~\cite{berthelot2018understanding}. Deep neural networks and domain knowledge were combined to improve the performance of the interpolation~\cite{FlowNet-2017-Ilg}.

\section{Background}
\label{sec:background}

% \my{Some of your related work should move in here. }

\subsection{4D Seismic Imaging and Sleipner Geology}

% \my{Should this be more relevant to Sleiner data, i.e., post-stack seismic imaging? You can also provide more details on Sleiner data set.}

4D post-stack seismic imaging for monitoring CO$_2$ sequestration involves repeated seismic surveys in time-lapse to image the subsurface changes due to CO$_2$ injection~\cite{lumley20104d} over time. It provides an understanding of the spatial movement and distribution of hydrocarbon and fluids, including supercritial CO$_2$ and the saline water saturated by CO$_2$, over time. The seismic data with an initial baseline survey before the injection and several follow-up surveys after a few years of injection are collected and processed with the same processing parameters. The anomalies between the processed data with the baseline survey and the follow-up surveys are attributable to changes in the reservoir~\cite{sambo2020role}, such as fluid movement~\cite{behrens20024d} and CO$_2$ injection~\cite{ivanova2012monitoring}.

\begin{figure}[t]
\centering
\includegraphics[width=1\columnwidth]{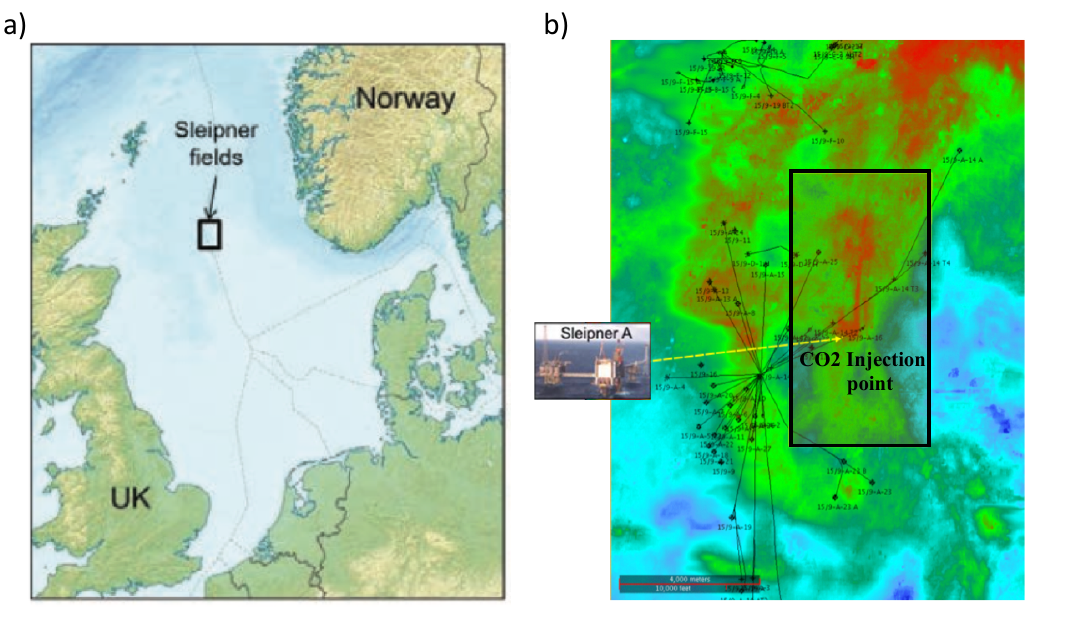}
\caption{ a) Location map~\cite{furre201720}. b) Utsira two-way traveltime (TWT) Map. Red is shallow, blue is deep~\cite{Utsira}.}
\label{fig:sleipner}
\end{figure}

The subsurface geology of the Sleipner field in the North Sea has been well studied and shown to be ideal for CO$_2$ sequestration~\cite{boait2012spatial}. Particularly, the storage reservoir is a saline aquifer of regional extent stretching for more than 400~km north to south and 50$\sim$100~km east to west, locally exceeding thicknesses of 300~m~\cite{zweigel2004reservoir} shown in Figure~\ref{fig:sleipner}. CO$_2$ is injected into highly porous Utsira Formation, a saline aquifer approximately 1~km below the surface and 200~km from the coast~\cite{cavanagh2014sleipner}. The injection is conducted through a single near-horizontal injection well at a depth of 1,012~m below sea level~(bsl)~\cite{eiken2019twenty}. 4D time-lapse seismic surveys were implemented from 1994 to 2016. One pre-injection dataset and ten post-injection datasets are collected in a 4~km $\times$ 7~km~area. The collected data are processed to generate 3D seismic stacked sections~\cite{feng2019zero}. With the injection of CO$_2$, a growing CO$_2$ plume is manifested as a series of bright reflections in the sections~\cite{boait2012spatial}. The reflections mostly arise from thin layers of CO$_2$ trapped beneath very thin intra-reservoir mudstone and the reservoir caprock~\cite{chadwick2019forensic}. Moreover, the plume also causes a velocity push-down since the seismic waves travel much slower through CO$_2$-saturated rock than through the virgin aquifer~\cite{chadwick2010quantitative}.

Repeated 3D seismic data were collected in 1994, 1999, 2001, 2002, 2004, 2006, 2008, 2010, 2013, and 2016 for monitoring. However, only data from 1999 to 2010 are openly released. The total volume of injected CO$_2$ in each year is shown in Table~\ref{tab:inject}. The dataset in each year is a 3D post-stack seismic section with 2 seconds recording in a time spacing of 0.002~s. Each section has 480 grids in the inline direction and 144 grids in the crossline direction. The data differences caused by the CO$_2$ injection are mostly after 0.5~s. We use a Tesla V100 GPU in the training of the networks. Using the full temporal length (i.e., from 0.5~s to 2.0~s) will, unfortunately, exceed the GPU memory limit. Via numerous trying and testing, we found out that an image data with a temporal length from 0.5~s to 1.56~s is the maximum allowable size that not only fits right in our GPU device, but also contains all the critical information~(i.e., the overburden, the new reflections generated by CO$_2$ plume, and the shift of the reflection events). As a result, we extract the seismic sections from 0.5 to 1.56~s, resulted in the size of the extracted seismic sections as 480$\times$480$\times$144. A total of 8 extracted seismic sections from 1994 to 2010 are available for this study.

\subsection{Autoencoders and Latent Representation}

The autoencoder is one of the prevalent unsupervized methods to generate a compressed representations of its input ~\cite{Bank-2020-Autoencoders}. The representation is usually exploited to discover the underlying structural patterns of training data in the low-dimensional latent space by downstream machine learning models.
In general, the autoencoder consists of two parts: an encoder network and a decoder network. The encoder network aims to map the data from high-dimensional input to a lower-dimensional latent space. The decoder network recovers the data from the latent space. The encoder and decoder networks are trained simultaneously to minimize reconstruction errors. Unlike conventional dimension reduction methods, e.g., principal component analysis~(PCA), it is a non-linear compression model transforming inputs to a lower-dimensional latent space, which can capture the complex distribution of training data. The compressed representation can be reconstructed back to the original data space that ensures that the representation includes the critical information of inputs.

\begin{figure*}[ht]
\centering
\includegraphics[width=1.7\columnwidth]{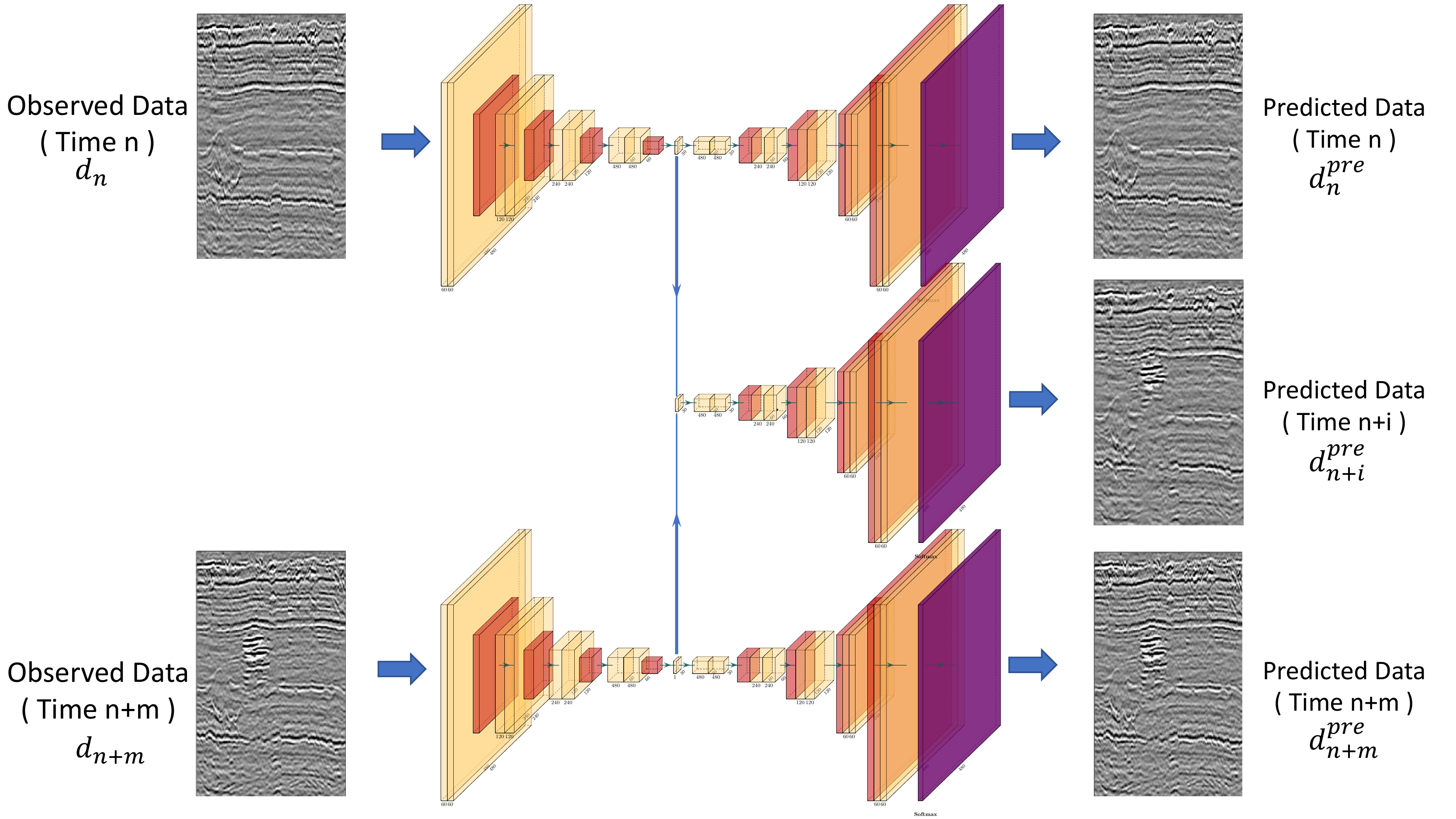}
\caption{A schematic illustration of the interpolation network which contains two encoder networks and three decoder networks. The observed data at time $n$ and $n+m$ are fed into the encoder networks to extract the latent spaces. Then the latent spaces and their linear combination are fed in the decoder networks to generate the data at $n$, $n+i$ and $n+m$.}
\label{fig:inte_net}
\end{figure*}

\subsection{Optical Flow}
Optical flow is a technique, originating in computer vision, to detect the motion of objects between consecutive frames of the sequence~\cite{li2020complete,zhang2014rtm}. It consists of vectors at each pixel location that represent the motions of the pixel from one frame to another. Particularly, given two successive images $P(x,y,z,t_1)$ and $P(x,y,z,t_2)$, we need to find the motion vector $\mathbf{m}(x,y,z)$ from $P(x,y,z,t_1)$ and $P(x,y,z,t_2)$. In 4D seismic imaging problem, $P(x,y,z,t_1)$ and $P(x,y,z,t_2)$ are seismic data at two close sequential times. A point at location $(x,y,z)$ in $P(x,y,z)$ will move by $\Delta{x}$, $\Delta{y}$ and $\Delta{z}$ after the time interval $\Delta{t}$ and this can be represented as:
\begin{equation}
P(x,y,z,t)=P(x+\Delta{x},y+\Delta{y},z+\Delta{z},t+\Delta{t}).
\label{eq:p_move}
\end{equation}
Assuming the movement to be small, the image at $P(x+\Delta{x},y+\Delta{y},z+\Delta{z},t+\Delta{t})$ can be expanded using a Taylor series:
\begin{eqnarray}
&P(x+\Delta{x},y+\Delta{y},z+\Delta{z},t+\Delta{t})=\nonumber\\
&P(x,y,z,t)+P_{x}\Delta{x}+P_{y}\Delta{y}+P_{z}\Delta{z}+P_{t}\Delta{t}+...,
\label{eq:p_taylor}
\end{eqnarray}
where $P_{x}$, $P_{y}$, $P_{z}$ and $P_{t}$ are partial derivative with respect to $x$, $y$, $z$ and $t$.
From Eqs.~(\ref{eq:p_move}) and (\ref{eq:p_taylor}), ignoring the higher order terms~(i.e.,``...'' in Eq.~(\ref{eq:p_taylor})), it follows that 
\begin{equation}
P_{x}u+P_{y}v+P_{z}w+P_t=0,
\label{eq:optical}
\end{equation}
where $u$, $v$ and $w$ are the $x$, $y$, $z$ components of the motion vector $\mathbf{m}(x,y,z)$, which are defined as
\begin{equation}
u=\frac{\Delta{x}}{\Delta{t}}, v=\frac{\Delta{y}}{\Delta{t}}, w=\frac{\Delta{z}}{\Delta{t}}.
\label{eq:motion}
\end{equation}
Combining the $\ell_2$ loss function with smoothing regularization, the objective function can be written as 
\begin{equation}
\zeta^\mathrm{op}=\iiint(P_{x}u+P_{y}v+P_{z}w+P_t)^2+\lambda^2\,M^2\,dxdydz,
\label{eq:op_loss}
\end{equation}
where $\lambda$ is the weight for the smoothing regularization, and $M$ is the magnitude of the flow gradient given by
\begin{equation}
M({u},{v},{w})=\sqrt{\|\nabla{u}\|^2+\|\nabla{v}\|^2+\|\nabla{w}\|^2}.
\label{eq:op_m}
\end{equation}

An iterative solution can be obtained as 
\begin{equation}
u^\mathrm{k+1}=\bar{u}^\mathrm{k}-\frac{P_{x}({P_{x}\bar{u}^\mathrm{k}+P_{y}\bar{v}^\mathrm{k}+P_{z}\bar{w}^\mathrm{k}+P_t})}{\alpha_{op}^2+P_{x}^2+P_{y}^2+P_{z}^2},
\label{eq:u}
\end{equation}
\begin{equation}
v^\mathrm{k+1}=\bar{v}^\mathrm{k}-\frac{P_{y}({P_{x}\bar{u}^\mathrm{k}+P_{y}\bar{v}^\mathrm{k}+P_{z}\bar{w}^\mathrm{k}+P_t})}{\alpha_{op}^2+P_{x}^2+P_{y}^2+P_{z}^2},
\label{eq:v}
\end{equation}
\begin{equation}
w^\mathrm{k+1}=\bar{w}^\mathrm{k}-\frac{P_{z}({P_{x}\bar{u}^\mathrm{k}+P_{y}\bar{v}^\mathrm{k}+P_{z}\bar{w}^\mathrm{k}+P_t})}{\alpha_{op}^2+P_{x}^2+P_{y}^2+P_{z}^2},
\label{eq:w}
\end{equation}
where $\bar{u}$, $\bar{v}$ and $\bar{w}$ are the local spatially smoothed values of $u$, $v$ and $w$, respectively. $k$ and $k+1$ are the iteration numbers and $\alpha_{op}$ is a hyper-parameter that helps avoid division by zero.

\section{Methods}

\begin{figure*}[ht]
\centering
\includegraphics[width=1.7\columnwidth]{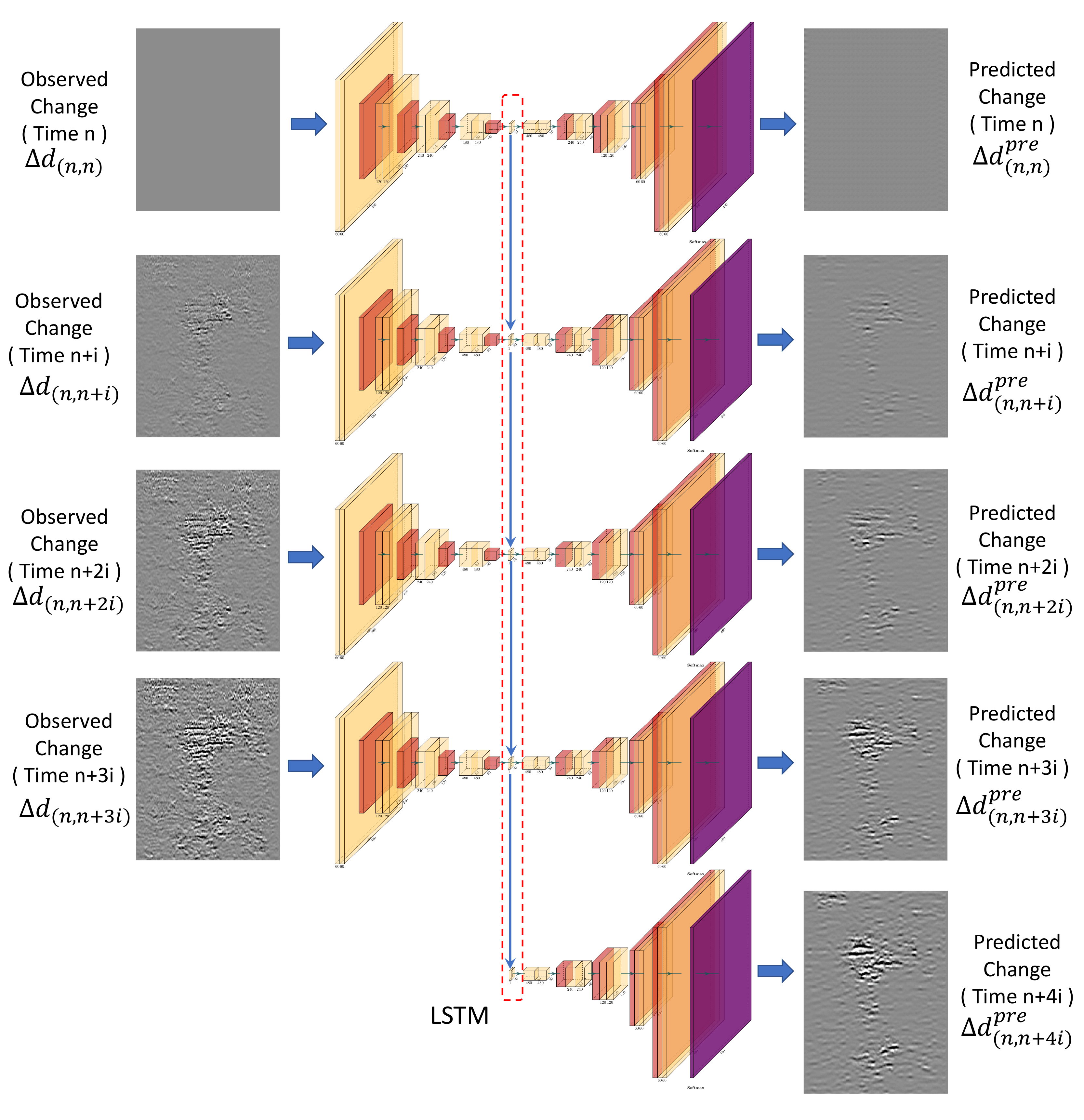}
\caption{A schematic illustration of the extrapolation network which contains four encoder networks and five decoder networks. The observed data change from time $n$ to $n+3i$ are fed into the encoder networks to extract the latent spaces. The latent spaces are connected using an LSTM network to predict the latent space at time $n+4i$. All these latent spaces are fed in the decoder networks to predict the data change from $n$ to $n+4i$.}
\label{fig:extra_net}
\end{figure*}

\subsection{Method Overview}

In this section, we develop two different networks for interpolation and extrapolation. For the interpolation task, inspired by the variational autoencoder~\cite{kingma2014auto}, we assume a linear pattern among samples in the latent space when the CO$_2$ volume increases over time. Thus, time sequences of samples can be generated by linearly interpolating features in the latent space. The extrapolation task is much more challenging in the sense that the problem is more severely ill-posed than interpolation. Prior knowledge needs to be exploited to make the problem well-posed so that reasonable solutions may be obtained. Here, we consider the consistency of the physics of the CO$_2$ migration throughout the monitoring, which leads us to incorporate the LSTM structure and the optical-flow regularization with our extrapolation network model. 

\subsection{Interpolation Network}

Due to the lack of sufficient observations of the underground CO$_{2}$ plume for training our interpolation network, it is desirable to introduce regularization during training to alleviate the overfitting risk. One reasonable regularization is to assume that the representations lie on the same line in the latent space. Although the migration patterns are actually nonlinear, our interpolation network can project time-series observations onto a line by nonlinear transformation with multiple convolutional layers. 

Our interpolation network consists of two encoder networks and three decoder networks, as shown in Figure~\ref{fig:inte_net}. Both the encoders and the decoders share their weights with other encoders and decoders. The observed seismic data with total CO$_2$ injected volume $n$ and $n+m$, $d_n$ and $d_{n+m}$, are fed to the encoders to obtain their latent space representations $L_{n}$ and $L_{n+m}$. Based on the linear regularization in the latent space,
%Assume the change of the latent space with the CO$_2$ injection %volume is linear, then 
the interpolated latent space $L$ with total CO$_2$ injected volume $n+i$ can be represented as
\begin{equation}
L_{n+i}=L_{n}+\frac{m-i}{m-n}(L_{n+m}-L_{n}),
 \label{eq:interpolate}
\end{equation}
where $m>i>0$. Both $L_{n}$, $L_{n+m}$ and the interpolated latent space $L_{n+i}$ are inserted in the decoders to construct the predicted seismic data with total CO$_2$ injected volume $n$, $n+i$ and $n+m$, ($d_{n}^\mathrm{pre}$, $d_{n+i}^\mathrm{pre}$ and $d_{n+m}^\mathrm{pre}$). The loss function of the network is defined as
\begin{equation}
\zeta^g=\sum_{j\in \{0,i,m\}}\sum_{n}\alpha^\mathrm{g}_{n+j}\left \|d_{n+j}-d_{n+j}^\mathrm{pre}\right \|^2,
\label{eq:loss}
\end{equation}
where $\alpha^\mathrm{g}$ are the weights for seismic data at different total CO$_2$ injected volume for the training of interpolation network.

\subsection{Extrapolation Network}

It is worth considering how extrapolation is even possible at all for our problems. The notion of extrapolation is the ability to make inferences that go beyond the scope of one's experiences. The generalization ability exhibited by contemporary neural network algorithms is largely limited to interpolation between data points in the training corpora so that the extrapolation can be a much challenging task for machine learning problems~\cite{Bishop-2006-Pattern}. However, fundamentally different from those problems from the computer vision community, there are unique physical laws underlying our problems. These physical laws are, to some extent, ``invariant'' in the sense that the spatial and temporal dynamics can be characterized by those laws throughout the physical process. Thus, the extrapolation task for our problems can be enabled by encouraging the learning of representations that reflect these ``invariant'' physical laws. It has been shown in different scientific applications that these physical laws can be effectively and implicitly captured via training neural networks with sufficient training data~~\cite{sun2020surrogate, Gomez-2020-Physics,feng2021superpixel}. 

Specific to our problems, with the interpolation network developed, we can use it to generate more simulations to augment the original dataset for training our extrapolation network. To better learn the complex temporal dynamics of CO$_2$ plume migration, we incorporate a long short-term memory~(LSTM) layer within our extrapolation network. The LSTM takes the historical latent representations as input and predicting the future states. An illustration of the extrapolation network is provided in Figure~\ref{fig:extra_net}. Unlike the interpolation network for generating the full seismic data~(shown in Figure~\ref{fig:inte_net}), the extrapolation network focuses on the change of the data. The extrapolation network is composed of four encoder networks and five decoder networks. Similar to the interpolation network, both the encoders and the decoders share their weights with other encoders and decoders. The seismic data with total CO$_2$ injected volume $n$, $d_{n}$ is set as the baseline and the change of the data with total CO$_2$ injected volume from $n$ to $n+3i$ ($\Delta{d_{(n,n)}}$, $\Delta{d_{(n,n+i)}}$, $\Delta{d_{(n,n+2i)}}$, $\Delta{d_{(n,n+3i)}}$) are calculated,
\begin{equation}
\Delta{d_{(n,n+i)}}=d_{n+i}-d_{n},
 \label{eq:data_diff}
\end{equation}
where $\Delta{d_{(n,n)}}$, $\Delta{d_{(n,n+i)}}$, $\Delta{d_{(n,n+2i)}}$, $\Delta{d_{(n,n+3i)}}$ are the inputs for the encoder networks, and their latent space representations $l_{n}$, $l_{n+i}$, $l_{n+2i}$ and $l_{n+3i}$ are connected using LSTM network to predicted the latent space representation of the change with total CO$_2$ injected volume $n+4i$, $l_{n+4i}$. Then $l_{n+4i}$ is fit in the decoder network to 
generate the data change $\Delta{d^\mathrm{pre}_{(n,n+4i)}}$. The loss function of our extrapolation network is 
\begin{equation}
 \zeta^e=\sum_{m=0}^4\sum_{n}\alpha^\mathrm{e}_{n+mi} \left \|\Delta{d_{(n,n+mi)}}-\Delta{d^\mathrm{pre}_{(n,n+mi)}} \right \|^2,
 \label{eq:loss_extra}
\end{equation}
where $\alpha^\mathrm{e}$ is the weight for seismic data with different total CO$_2$ injected volume for the training of extrapolation network. In the last step, the predicted data change $\Delta{d^\mathrm{pre}_{(n,n+4i)}}$ is added to the baseline $d_{n}$ to generate the predicted data at time $n+4i$, $d^\mathrm{pre}_{n+4i}$.

\subsection{Extrapolation Network with Optical Flow Regularization}

As pointed out in \cite{boait2012spatial}, a prominent physical phenomenon that can be observed during the CO$_2$ injection at Sleipner is the ``velocity push-down'' effect. It is caused by the decrease of the velocity within the CO$_2$ reservoir after the injection of CO$_2$ in saturating local saline water. An immediate impact of this saturation would result in a longer traveltime for those seismic waves that propagate through the CO$_2$ reservoir. That in turn will result in a time-shift, (moving downward to be more specific), in the time-domain seismic image. During the process of CO$_2$ injection, the velocity of the CO$_2$ reservoir will keep decreasing due to the increase of the saturation of CO$_2$ within saline water. This will result in a continuous moving-downward of the reflection events below the CO$_2$ reservoir, which will be consistent throughout the injection process. That would provide a reliable constraint for extrapolation.  
%This unique physical pattern can be used to constrain our model. %Specifically, we use optical flow to detect the shift of the seismic response and set it as a regularization term in the training of extrapolation network. 

The technique of optical flow in Section III.C, as a method to describe the spatial movement of an object, can both capture the migration of CO$_2$ plume and the downward shifting of the reflection events. Given the seismic data with total CO$_2$ injected volume $n$, and $n+3i$, ($d_n$, $d_{n+3i}$), we calculate the observed optical flow vector $\mathbf{m_1}=(u_{1},v_1,w_1)$ from $d_n$ to $d_{n+3i}$. Given the predicted data with total CO$_2$ injected volume $n$, $n+i$ and $n+4i$, ($d_n$, $d_{n+i}$, $d_{n+4i}$), we calculate the predicted vector $\mathbf{m_2}=(u_2,v_2,w_2)$ from $d^\mathrm{pre}_{n}$ to $d^\mathrm{pre}_{n+4i}$ and the predicted vector $\mathbf{m_3}=(u_3,v_3,w_3)$ from $d^\mathrm{pre}_{n+i}$ to $d^\mathrm{pre}_{n+4i}$. 
The regularization term is defined as the summation of the included angle between $\mathbf{m_1}$ and $\mathbf{m_2}$ and the included angle between $\mathbf{m_1}$ and $\mathbf{m_3}$ ,
\begin{equation}
\zeta^\mathrm{reg}=\sum_{n}\left \|\arccos{r}_n+\arccos{r}_{n+1}\right \|^2,
\label{eq:loss_reg}
\end{equation}
where $\arccos$ is the inverse of the cosine function,
\begin{equation}
{r}_n=\frac{u_{1}u_{2}+v_{1}v_{2}+w_{1}w_{2}}{\sqrt{u_1^2+v_1^2+w_1^2}\sqrt{u_2^2+v_2^2+w_2^2}},
\label{eq:angle_n1}
\end{equation}
and
\begin{equation}
{r}_{n+1}=\frac{u_{1}u_{3}+v_{1}v_{3}+w_{1}w_{3}}{\sqrt{u_1^2+v_1^2+w_1^2}\sqrt{u_3^2+v_3^2+w_3^2}}.
\label{eq:angle_n2}
\end{equation}
Thus, the overall loss function for the extrapolation network with optical flow regularization can be posed as
\begin{equation}
\zeta^\mathrm{e_{reg}}=\zeta^\mathrm{e}+\alpha_{reg}\,\zeta^\mathrm{reg},
\label{eq:loss_e_reg}
\end{equation}
where the terms of $\zeta^\mathrm{e}$ and $\zeta^\mathrm{reg}$ are provided in Eqs.~\eqref{eq:loss_extra} and \eqref{eq:loss_reg}, respectively, and $\alpha_{reg}$ is the weight for the regularization term. It is worth mentioning that this regularization only constrains the direction of the seismic data change, not the amplitude of the seismic data change.

\begin{figure*}[ht]
\centering
\includegraphics[width=2\columnwidth]{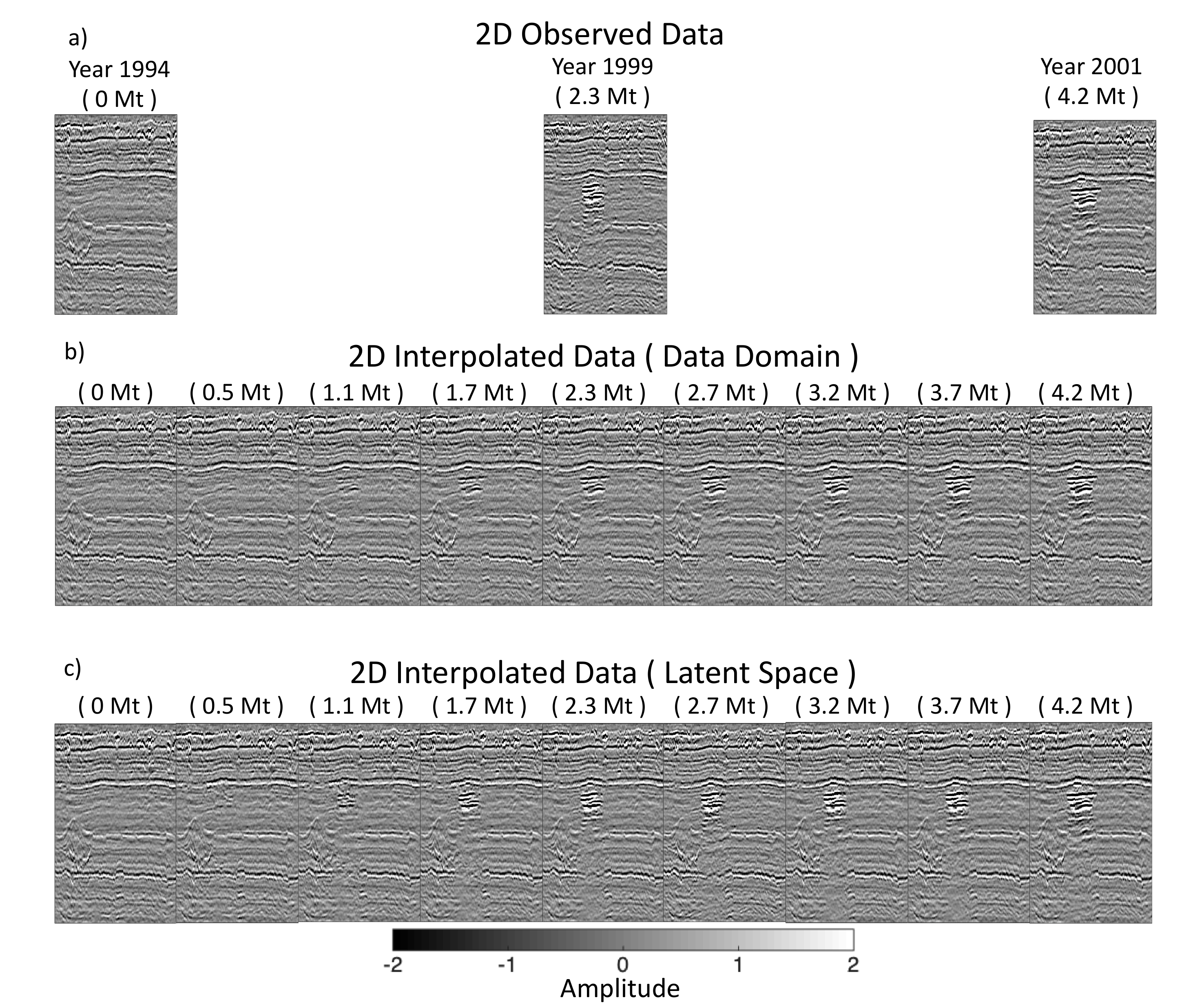}
\caption{ a) 2D observed data, interpolated data in the b) data domain and c) latent space from year 1994 to 2001. For the data domain interpolation, the interpolated data are generated by linear combination of the observed data. For the interpolation in the latent space, the latent spaces of the observed data are combined linearly and then fed into the decoders to generate interpolated data.}
\label{fig:inte_data}
\end{figure*}
\begin{figure*}[ht]
\centering
\includegraphics[width=2\columnwidth]{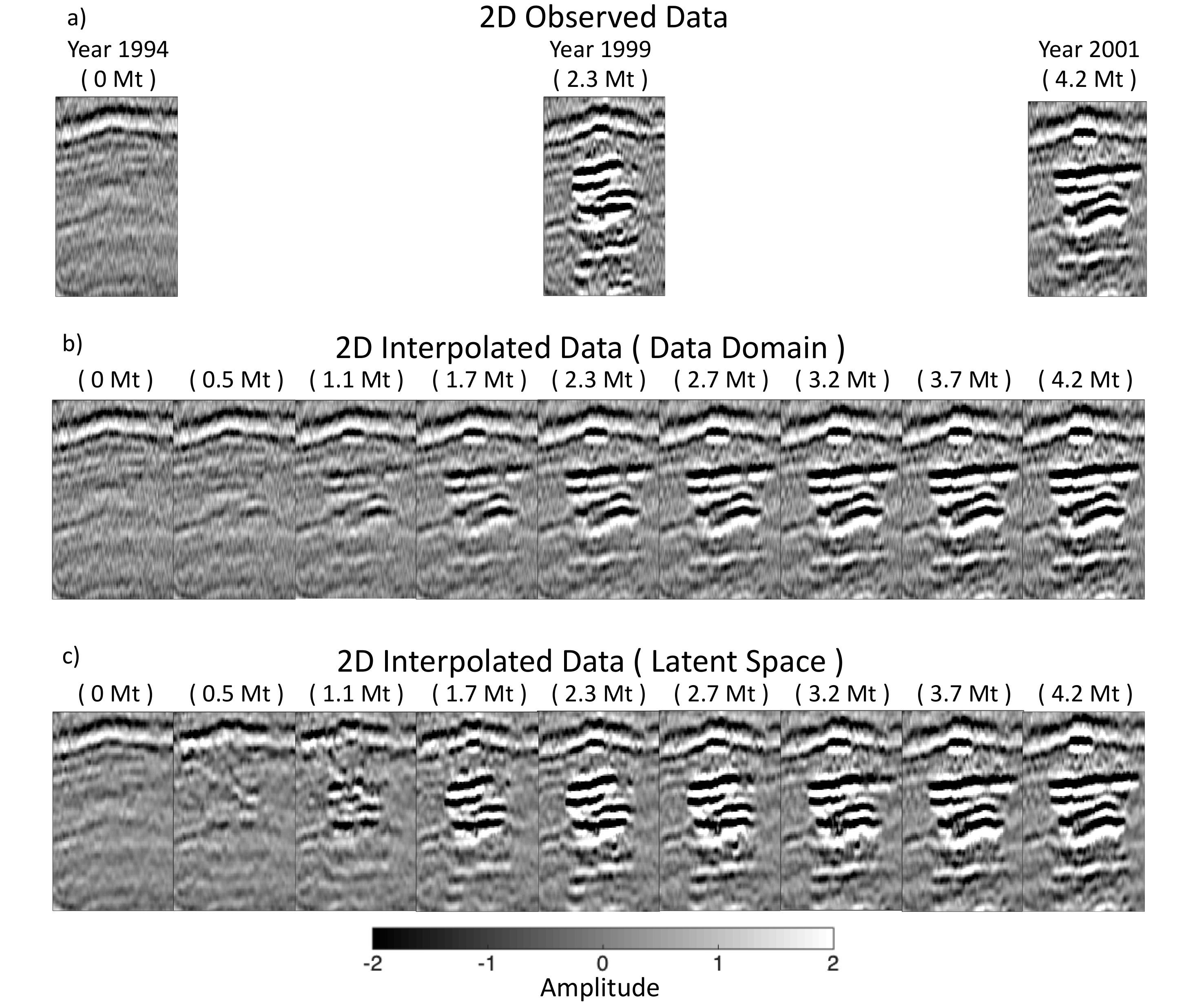}
\caption{ The zoom views of a) 2D observed data, interpolated data in the b) data domain and c) latent space from year 1994 to 2001.}
\label{fig:inte_data2}
\end{figure*}

\section{Tests}
% \my{More details needed for Sleipner data and acquisition~(such as dimension, injection volume, some geologic info, etc.) This may go to Section III, Background.}

In this section, we provide numerical tests and results showing the performance of our models in both 2D and 3D interpolation~(Sections V.A \& V.B) and extrapolation~(Sections V.C \& V.D) scenarios. A side-by-side comparison to two baseline models (linear model and neural network model without regularization) is also provided.

\subsection{2D Sleipner Interpolation}

To test the 2D interpolation method, we select a 2D slice in the 3D seismic section and collect the seismic data of this slice every monitoring year. Among all acquired real seismic data over 8 years, a random selection of 3 subsets out of 8 existing datasets results in a total of 56 combinations (i.e., $C_8^3$). Each subset can be used as a training sample for our interpolation model in Figure~\ref{fig:inte_net}. In particular, three sequential seismic data are selected and used as the inputs (i.e., $d_n$, $d_{n+i}$ and $d_{n+m}$). We use 45 of the subsets as the training set and the rest are held out for validation. The weights in Eq.~\eqref{eq:loss} ( $\alpha^\mathrm{g}_{n}$, $\alpha^\mathrm{g}_{n+i}$ and $\alpha^\mathrm{g}_{n+m}$) are set as (0.2, 0.6, 0.2).

After 1,200 epochs of training, the mean-square-error~(MSE) loss value decreases to 0.06 for the training set and 0.09 for the validation set. An example of the interpolated data is shown in Figure~\ref{fig:inte_data}, where we provide three real seismic datasets as the ground truth for comparison. We also provide the synthetic seismic data generated using linear interpolation based on pixels~(namely, data domain), i.e., 
\begin{equation}
d_{n+i}=d_{n}+\frac{I_{n+m}-I_{n+i}}{I_{n+m}-I_{n}}*(d_{n+m}-d_{n}).
 \label{eq:interpolate_data}
\end{equation}
With the high fidelity synthetic seismic data being generated using our model, we are able to ``fill the gap'' between each year when no real seismic data are acquired. By looking at the zoomed view of the interpolated data in Figure~\ref{fig:inte_data2}c, we can see the expansion of reflectors caused by the growth of CO$_2$ plume. In comparison, only the cross-dissolve transition can be seen with the interpolation in the data domain in Figure~\ref{fig:inte_data2}b.

\begin{figure*}[ht]
\centering
\includegraphics[width=1.8\columnwidth]{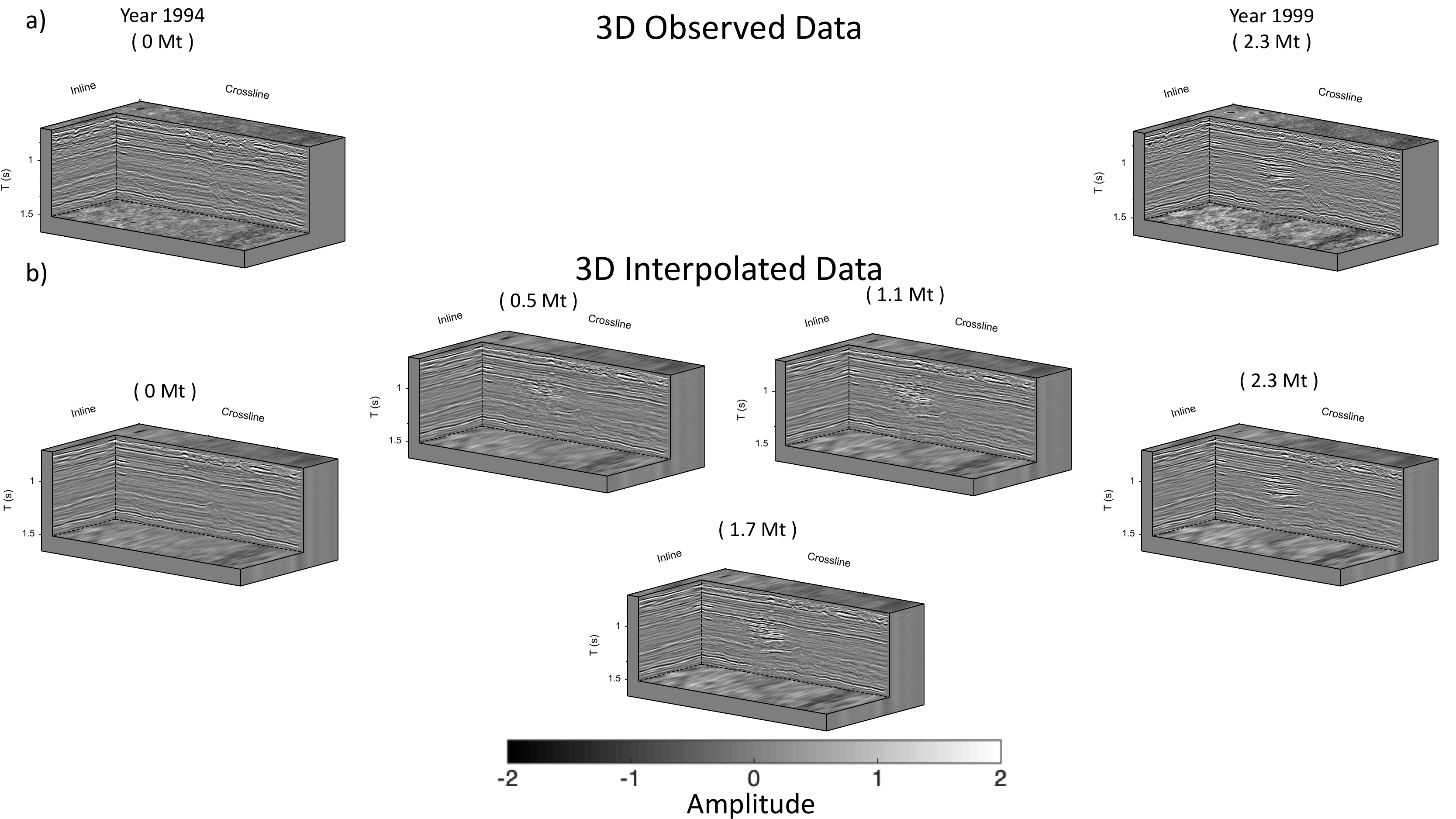}
\caption{ a) 3D observed data and b) interpolated data in the latent space from year 1994 to 1999.}
\label{fig:inte_data3d}
\end{figure*}

\begin{figure*}[ht]
\centering
\includegraphics[width=1.8\columnwidth]{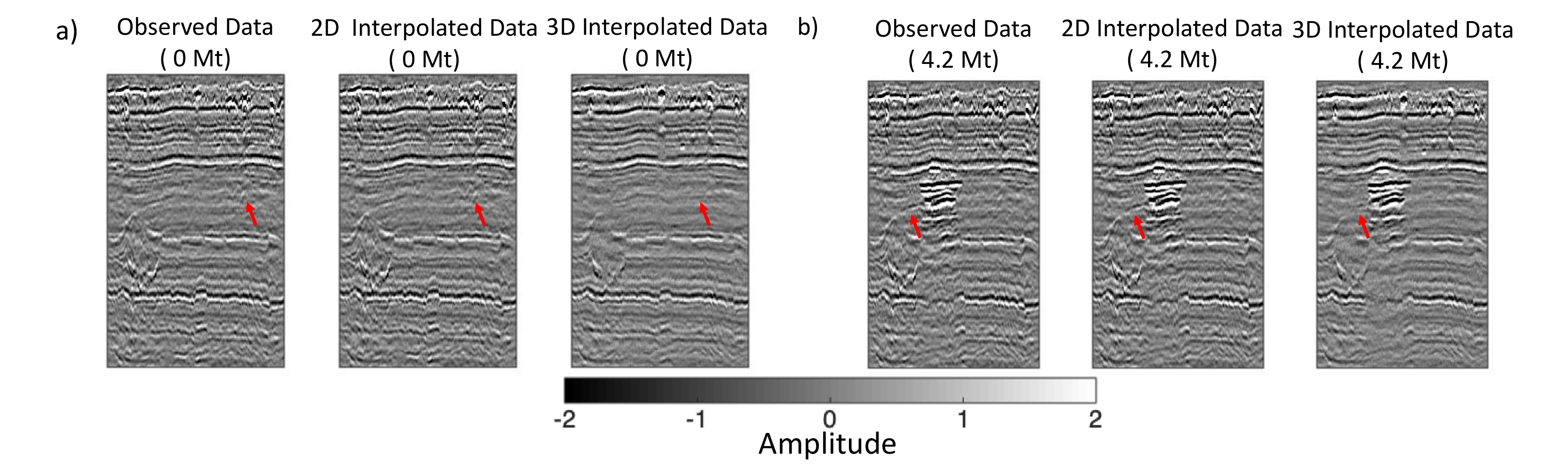}
\caption{ The comparisons of the observed data, 2D interpolated data, 3D interpolated data with injection volume of a) 0 Mt and b) 4.2 Mt. }
\label{fig:inte_data_compare}
\end{figure*}

\subsection{3D Sleipner Interpolation}
We slice the 3D Sleipner data in a group of 2D seismic images along the crossline direction, where each 2D seismic image can be regarded as a channel image.
Since there are 144 grids in the crossline direction of the 3D Sleipner data, we have 144 channels for the input of the interpolation network. Due to the limitation of the computational cost, 2D convolution filters are used instead of the 3D convolution filters. The interpolation network structure is the same as the 2D case except that the number of the channels in the input and output layers are changed.  

After 1,200 epochs, the MSE loss decreases to 0.35 for the training set and 0.45 for the validation set, both of them are higher than the losses in the 2D case. Examples of interpolated data are shown in Figure~\ref{fig:inte_data3d} and comparisons between the observed data, 2D interpolated data and 3D interpolated data are shown in Figure~\ref{fig:inte_data_compare}, where some details are missing in the 3D interpolated data that the red arrows point. It means the results are affected by the computational cost problem. To achieve similar accuracy as the 2D case, 3D convolution filter, more training epochs and a larger neural network with more parameters are required.

%some details are lost in the 3D interpolated data where the red arrows point at.

%The data difference in the 3D interpolation is higher than the data is the difference in 2D. The MSE and SSIM
%

\begin{figure}[ht]
\centering
\includegraphics[width=1\columnwidth]{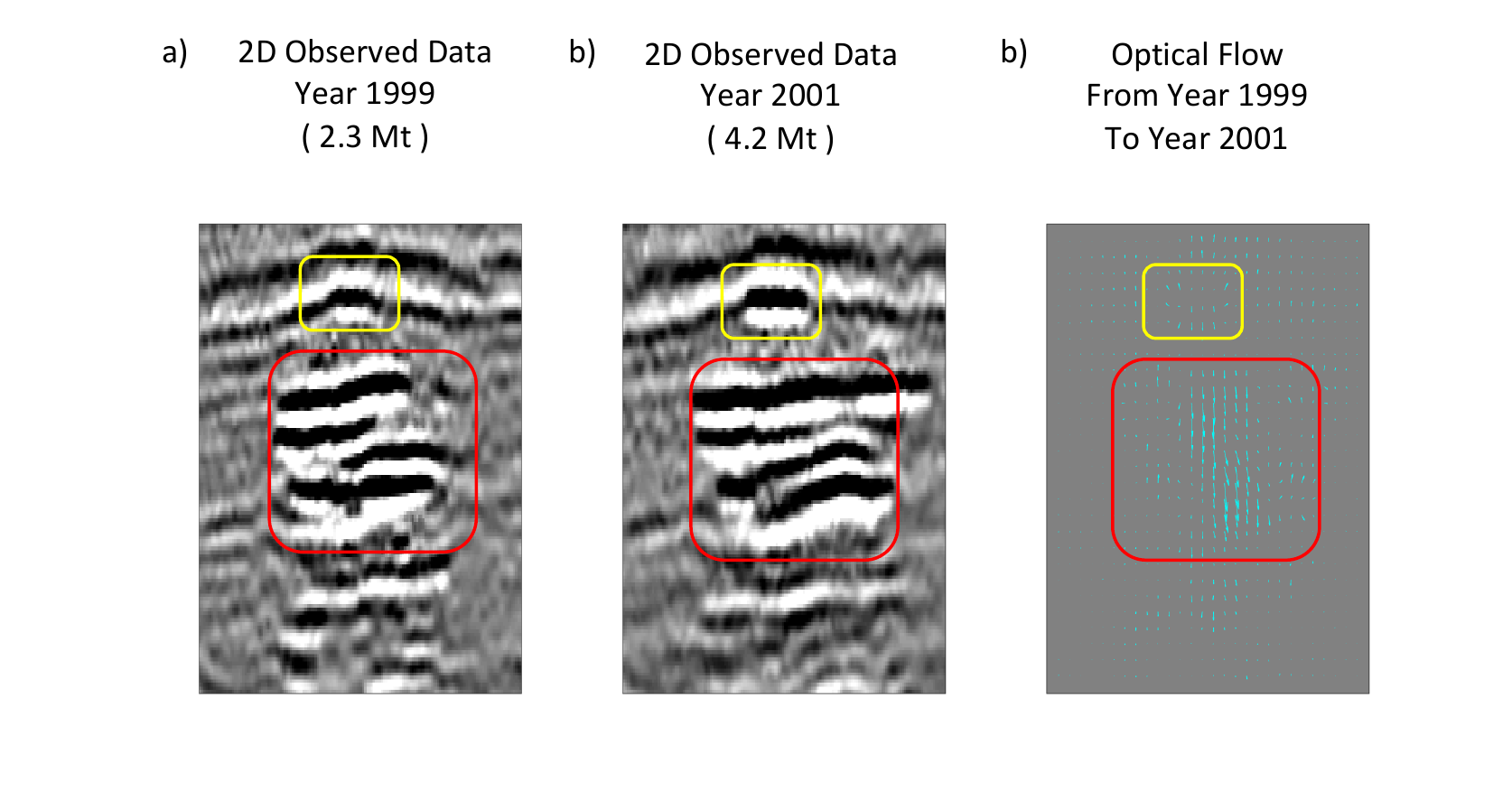}
\caption{ An example of optical flow: a) the 2D observed data in 1999, b) the 2D observed data in 2001 and c) the calculated optical flow using the data from 1999 to 2001. }
\label{fig:optical_example}
\end{figure}

\begin{figure*}[ht]
\centering
\includegraphics[width=1.4\columnwidth]{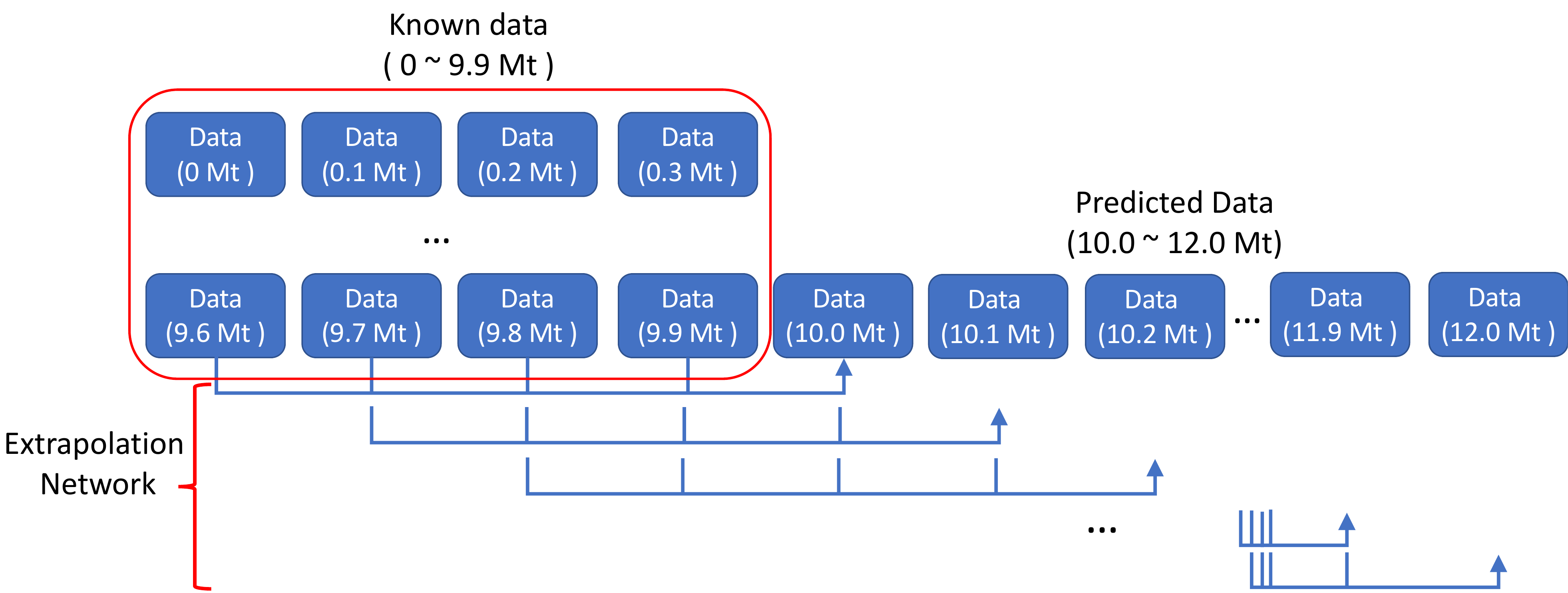}
\caption{ Workflow to predict data between 10.0 and 12.0 Mt from data 0 between 9.9 Mt using extrapolation network }
\label{fig:extra_flow}
\end{figure*}

\begin{figure*}[ht]
\centering
\includegraphics[width=1.4\columnwidth]{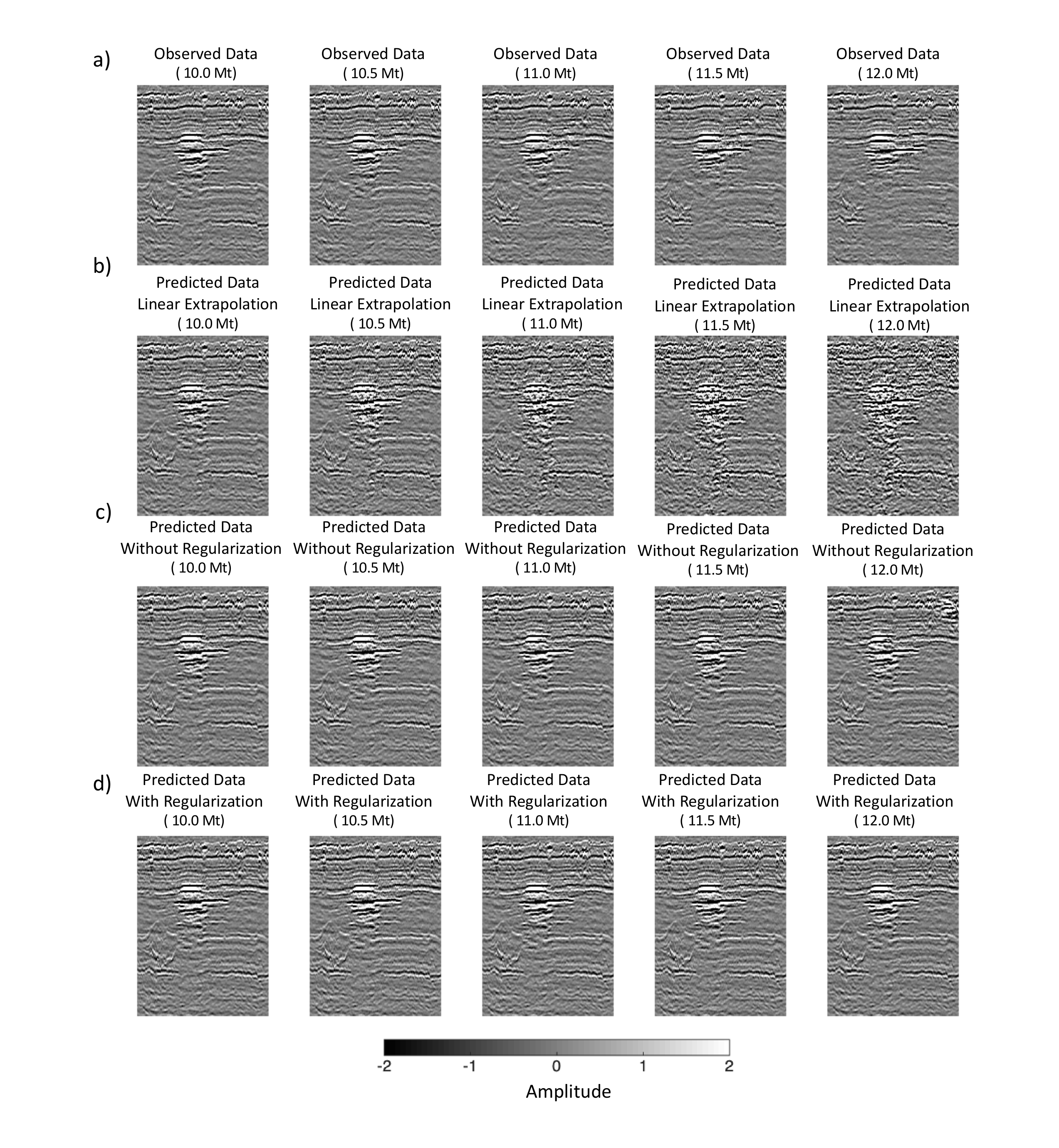}
\caption{ a) The observed data, b) by linear extrapolation, by extrapolation network c) without regularization and d) with regularization from 9.6 to 12.0 Mt  }
\label{fig:extra_full}
\end{figure*}

\begin{figure*}[ht]
\centering
\includegraphics[width=1.4\columnwidth]{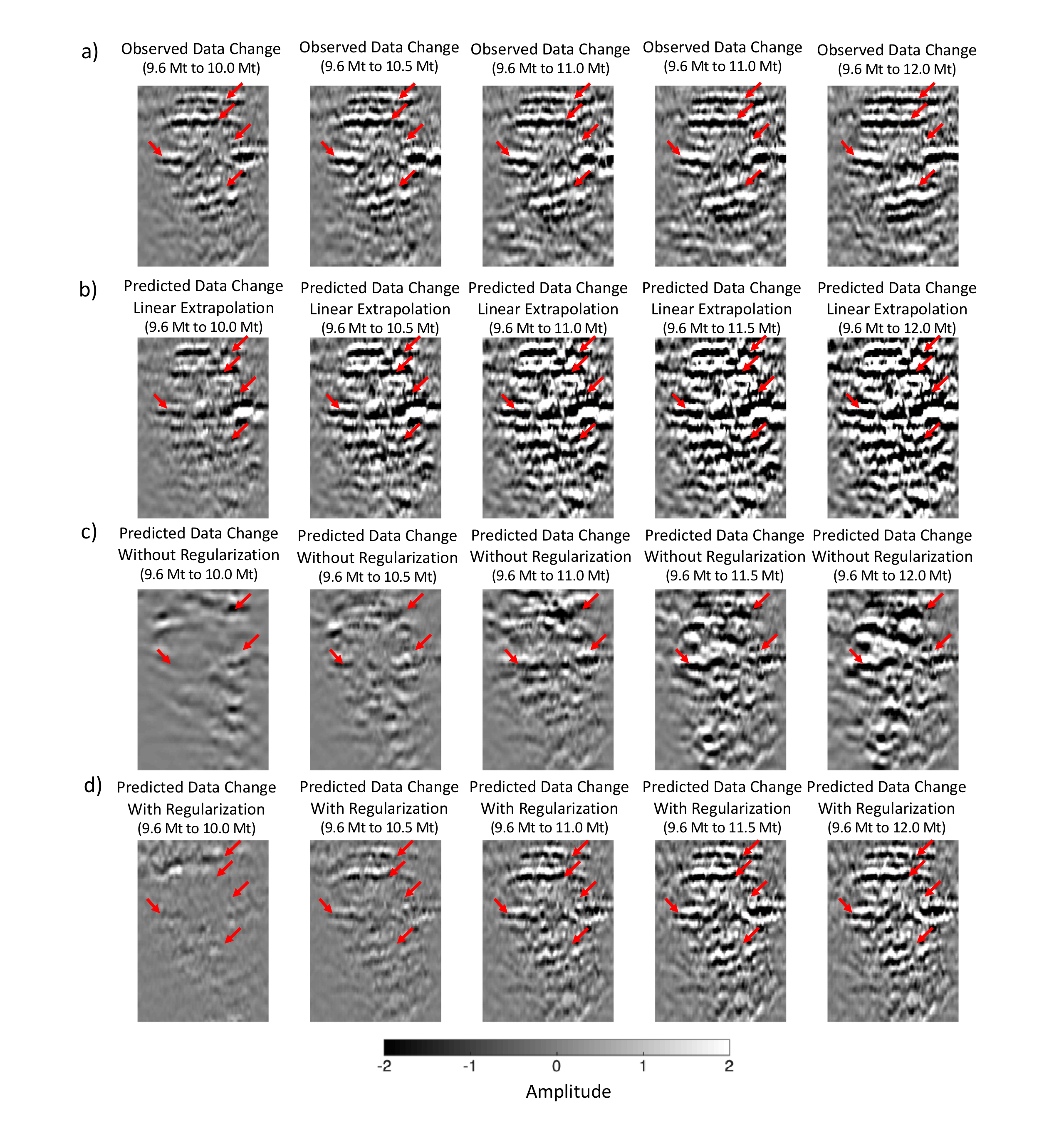}
\caption{ The zoom view of a) observed data change, predicted data change b) by linear extrapolation, by extrapolation network c) without regularization and d) with regularization from 9.6 to 12.0 Mt }
\label{fig:extra_change}
\end{figure*}

\subsection{2D Sleipner Extrapolation}
Only eight seismic datasets were collected in the 16 years of monitoring in the Sleipner area which is not enough for the training of the extrapolation network. To augment the dataset, we use the interpolation network to generate 121 seismic datasets every 0.1~Mt from 0 to 12~Mt. We assume that only the data from 0 to 9.9 Mt are known and these data are used in the training of the extrapolation network. The data from 10.0 to 12.0~Mt are used as the validation sets. The change of total CO$_2$ injected volume with each step in the extrapolation network, $i$, is set as 0.1~Mt.

During training, we set [$\alpha^\mathrm{e}_n$, $\alpha^\mathrm{e}_{n+i}$, $\alpha^\mathrm{e}_{n+2i}$, $\alpha^\mathrm{e}_{n+3i}$, $\alpha^\mathrm{e}_{n+4i}$] as (0.4, 0.4, 0.4, 0.4, 1) and the optical flow weight $\alpha_{reg}$ as 0.1. 2D optical flow is applied to calculate the shift of the seismic data caused by the injection of CO$_2$. An example is shown in Figure~\ref{fig:optical_example}. The expanding of the reflection event in the yellow box and the ``push-down'' effect of the events in the red boxes are clearly detected by the optical flow.

 After training for 1,200 epochs, the training loss decreases to 0.31 and the validation loss decreases to 0.50. The workflow to predict data between 10.0 and 12.0 Mt from data 0 between 9.9 Mt is shown in Figure~\ref{fig:extra_flow}. The data from 9.6 to 9.9 Mt are fed into the extrapolation network to predict the data with 10.0 Mt. Then the predicted data with 10 Mt is combined with the data from 9.7 to 9.9 Mt to predict the data with 10.1 Mt and the rest can be done in the same manner. As the injection volume increases, the predicted data will move further away from the data domain we know from 0 to 9.9 Mt. As a result, the data from the prediction will be more and more unreliable as the injection volume increases. The predicted data from 10.0 to 12.0 Mt is shown in Figure~\ref{fig:extra_full} and the MSE and the mean absolute error (MAE) are shown in Table~\ref{tab:with_reg}, the MSE and MAE losses increase with the increasing of the total CO$_2$ injected volume. 

Since the velocity change is small with a large amount of CO$_2$ trapped in the reservoir, the change in the seismic data is subtle when the CO$_2$ injected volume is over 10 Mt as shown in Figure~\ref{fig:extra_full}. Figure~\ref{fig:extra_change} shows the zoom view of the data change in the CO$_2$ plume area from 9.6 to 12.0 Mt. The predicted data generated by linear extrapolation in the data domain, extrapolation network without and with optical flow regularization are compared. There are five bright reflection events where the red arrows point in the real data. Their amplitudes increase with the increase of total CO$_2$ injected volume and the changing rate varies with time. For the data generated by linear extrapolation, the amplitudes of the events increase at a constant rate. As a result, the data change becomes very large when the total injection volume is 12 Mt in Figure~\ref{fig:extra_change}b, which is far away from the real data. In Figure~\ref{fig:extra_change}c, the predicted data change without regularization only detects three events where the arrows point. The predicted data change is distorted when the total injection volume is larger than 11.5 Mt. For the predicted data with regularization in Figure~\ref{fig:extra_change}d, all five events are predicted correctly. The trend of the increasing amplitude and expanding of the events remains the same even when the data is far away from the known data domain. As a result, both the MSE and MAE losses of the predicted data with regularization are less than the losses of the data without regularization and the data with linear extrapolation in table~\ref{tab:with_reg}.

\subsection{3D Sleipner Extrapolation}
For the 3D Sleipner data, we use the interpolation network to generate 24 seismic datasets every 0.5 Mt from 0 to 12 Mt. The data from 0 to 9.5 Mt are used to the train the extrapolation network. The data from 10.0 to 12.0 Mt are used as the validation set. The change of total CO$_2$ injected volume with each step in the extrapolation network, $i$, is 0.5 Mt. The weights [$\alpha^\mathrm{e}_n$, $\alpha^\mathrm{e}_{n+i}$, $\alpha^\mathrm{e}_{n+2i}$, $\alpha^\mathrm{e}_{n+3i}$, $\alpha^\mathrm{e}_{n+4i}$] are (0.4, 0.4 ,0.4, 0.4, 1) and the optical flow weight $\alpha_{reg}$ is 0.1.

\begin{figure*}[ht]
\centering
\includegraphics[width=1.6\columnwidth]{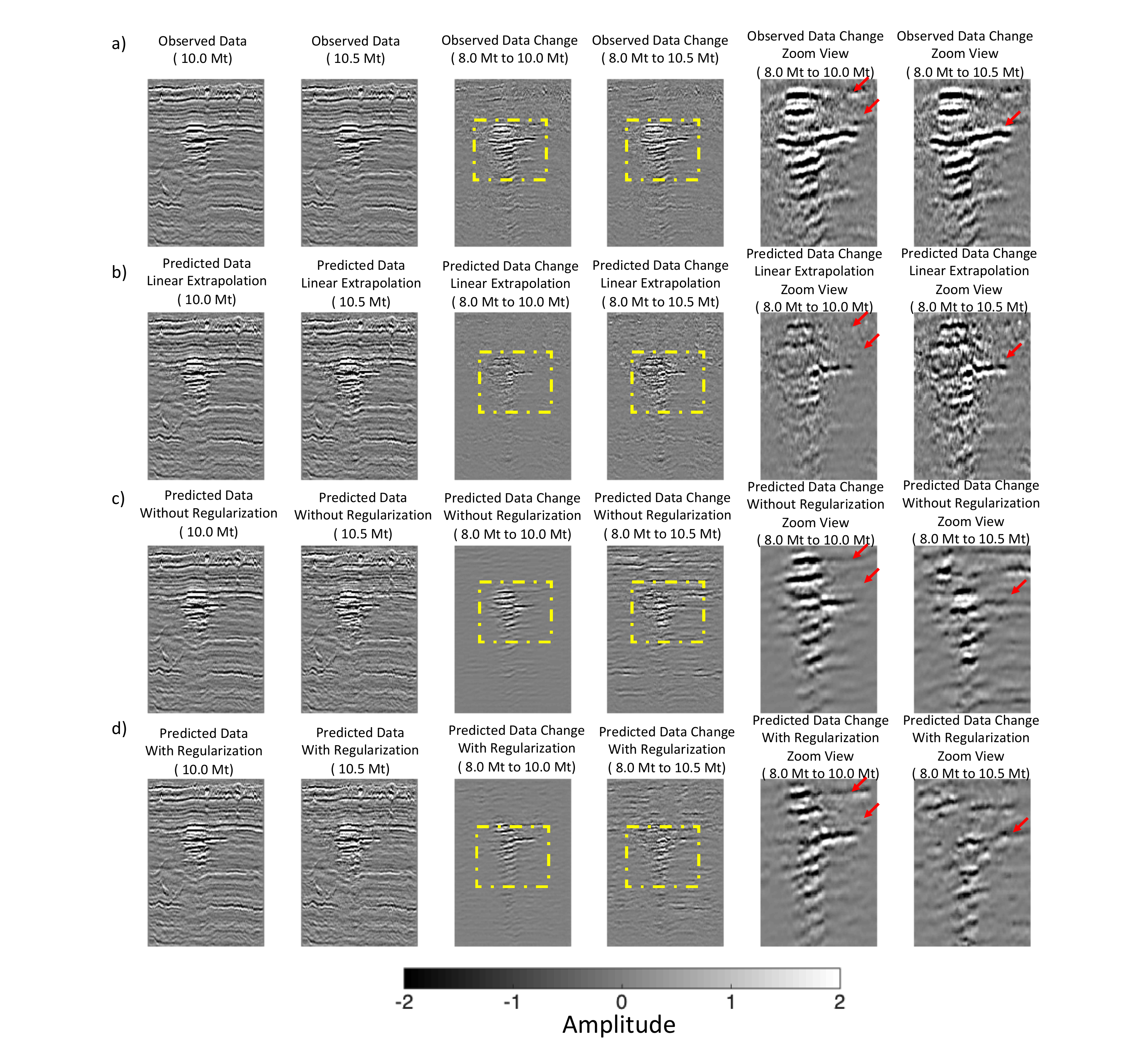}
\caption{ 2D slices of the 3D data, the data change and the data change zoom view in the yellow box of a) observed data, predicted data by b) linear extrapolation, c) without regularization and d) with regularization from 8.0 to 10.5 Mt.}
\label{fig:extra_3d}
\end{figure*}

After 600 epochs, the training loss decreases to 0.65 and the validation loss decreases to 0.83. We use the data from 8.0 to 9.5 Mt to predict the data from 10.0 to 12.0 Mt. The losses between the 3D extrapolated data and the true data are given in the table~\ref{tab:with_reg3D}. Similar to the 2D extrapolated data, the losses increase with the total CO$_2$ injected volume. Moreover, they increase much faster than the 2D cases since the change of total CO$_2$ injected volume in each step is much bigger. Examples of the slices in the 3D extrapolated data from 10.0 Mt to 10.5 Mt and the data change from 8.0 Mt to 10.5 Mt are shown in Figure~\ref{fig:extra_3d}. By looking at the areas that the red arrows points in the data change, we can see the data change generated by the extrapolation network with regularization is closer to the true data than the predicted data change generated by the extrapolation network without regularization.  Although the reflection events are preserved in for the predicted data change generated linear extrapolation, cycle-skipping~\cite{chen2020multiscale} occurs with these events due to the ``velocity push-down'' effect, which causes large MSE and MAE losses. The MSE and MAE losses of the predicted data are given in Table~\ref{tab:with_reg3D}, in which the data change predicted by the extrapolation network with regularization has the lowest losses.

\begin{table*}
\centering

\caption{The errors in MSE and MAE between the 2D extrapolated data and observed data}
\begin{tabular}{ccccccc} 
\hline
\multicolumn{1}{l}{}                                                                                               & ~CO$_2$ Injected Volume & ~10.0 Mt~ & ~10.5 Mt~ & ~11.0 Mt~ & ~11.5 Mt~ & ~12.0 Mt~  \\ 
\hline
\multirow{2}{*}{\begin{tabular}[c]{@{}c@{}}\textbf{Predicted Data}\\\textbf{Linear Extrapolation} \end{tabular}} & MSE                   & 0.1419    & 0.4994    & 1.1370    & 1.9074    & 2.8454     \\
                                                                                                                   & MAE                  & 0.1594    & 0.3191    & 0.5307    & 0.7177    & 0.8843     \\ 

\hline
\multirow{2}{*}{\begin{tabular}[c]{@{}c@{}}\textbf{Predicted Data}\\\textbf{Without Regularization} \end{tabular}} & MSE                   & 0.0529    & 0.1005    & 0.1562    & 0.1725    & 0.2178     \\
                                                                                                                   & MAE                  & 0.1486    & 0.2023    & 0.2751    & 0.3420    & 0.4229     \\ 
\hline
\multirow{2}{*}{\begin{tabular}[c]{@{}c@{}}\textbf{Predicted Data}\\\textbf{With Regularization} \end{tabular}}    & MSE                   & \textbf{0.0488}    & \textbf{0.0941}    & \textbf{0.1351}    & \textbf{0.1537}    & \textbf{0.1621}     \\
                                                                                                                   & MAE                  & \textbf{0.1475}    & \textbf{0.1970}    & \textbf{0.2569}    & \textbf{0.3085}    & \textbf{0.3509}     \\
\hline
\label{tab:with_reg}
\end{tabular}
\end{table*}

\begin{table*}
\centering

\caption{The errors in MSE and MAE between the 3D extrapolated data and observed data}
\begin{tabular}{ccccccc} 
\hline
\multicolumn{1}{l}{}                                                                                               & ~CO$_2$ Injected Volume & ~10.0 Mt~ & ~10.5 Mt~ & ~11.0 Mt~ & ~11.5 Mt~ & ~12.0 Mt~  \\ 
\hline
\multirow{2}{*}{\begin{tabular}[c]{@{}c@{}}\textbf{Predicted Data}\\\textbf{Linear Extrapolation} \end{tabular}} & MSE                   & 0.0475    & 0.2065    & 0.5200    & 0.9755    & 1.4381     \\
                                                                                                                   & MAE                  & 0.1079    & 0.2322    & 0.3751    & 0.5251    & 0.6421     \\ 
\hline
\multirow{2}{*}{\begin{tabular}[c]{@{}c@{}}\textbf{Predicted Data }\\\textbf{Without Regularization} \end{tabular}} & MSE                   & 0.1028    & 0.2064    & 0.5095    & 0.7933    & 1.1066     \\
                                                                                                                   & MAE                  & 0.1500    & 0.2589    & 0.4202    & 0.5418    & 0.6231     \\ 
\hline
\multirow{2}{*}{\begin{tabular}[c]{@{}c@{}}\textbf{Predicted Data }\\\textbf{With Regularization} \end{tabular}}    & MSE                   & \textbf{0.0925}    & \textbf{0.1338}    & \textbf{0.3089}    & \textbf{0.5079}    & \textbf{1.0611}     \\
                                                                                                                   & MAE                  & \textbf{0.1521}    & \textbf{0.2152}    & \textbf{0.3353}    & \textbf{0.4401}    & \textbf{0.6087}     \\
\hline
\label{tab:with_reg3D}
\end{tabular}
\end{table*}

\section{Discussions and Future Work}

\begin{figure*}[ht]
\centering
\includegraphics[width=1.7\columnwidth]{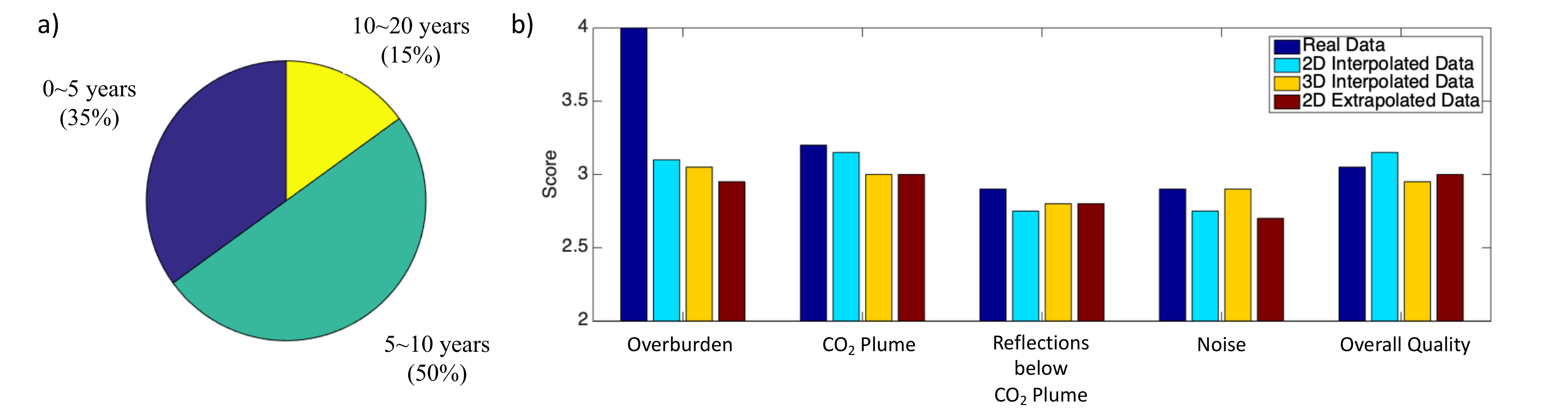}
\caption{ a) Working years of the experts in the seismic imaging area. b) Average scores of the expert evaluation on the quality of the data.}
\label{fig:evaluation}
\end{figure*}

\noindent \textbf{Expert Evaluation.~}Two error metrics~(i.e., MSE and MAE) have been used to evaluate the quality of our models in Tables~\ref{tab:with_reg} and \ref{tab:with_reg3D}. They are both pixel-based, which may be insufficient for assessing structured data such as images~\cite{Zhang-2018-Unreasonable}. To overcome potential biased assessment based on MSE and MAE metrics, we provide an expert evaluation, which will add another dimension of assessment. This is particularly important for our work since the resulting images will eventually be presented to the field operators for monitoring purposes. From that perspective, the standard for the evaluation would be to pass the domain expert with our generated samples. To rule out any subjective evaluation, we conduct a blind survey completed by 20 domain experts to evaluate the quality of our interpolation and extrapolation data. The participants' number of working years in the seismic imaging area are listed in Figure~\ref{fig:evaluation}a, which can be used as an indication of their level of expertise. The survey can be found in our online Google form~\footnote{\url{https://docs.google.com/forms/d/e/1FAIpQLSfEZVvuXHcIub_mVzmLGEIOA4ZXkFWL_Tz8sXwwG1nqRrcukA/viewform}}. In this survey, we design five questions regarding specific geologic features of overburden, CO$_2$ plume, reflection below CO$_2$ plume, noise, and overall quality. We pick those features since they have been utilized to characterize CO$_2$ migration in various work~\cite{CO2-2014-Chadwick, boait2012spatial}. Four different seismic images belonging to four different datasets are used in the survey including, real data, 2D interpolated data, 3D interpolated data, and 2D extrapolated data. We then ask the experts to grade (on a scale of 4 points) the quality of the images based on the aforementioned geologic features without telling the participants in advance whether the data is generated or not. Average scores of the expert evaluation are shown in Figure~\ref{fig:evaluation}b. 

For the overburden area, the score of true data is 4, which is much higher than the scores of 2D interpolated data, 3D interpolated data and 2D extrapolated data. This indicates the overburden area of the generated data needs further improvements. However, it is interesting to notice that for the rest of the geologic features, the scores are all very close. For example, the scores for the overall quality of the real data, 2D interpolated data, 3D interpolated data and 2D extrapolated data are [3.05, 3.15, 2.95, 3.00], which means the quality of the generated data is close enough to the real data. Even the experts cannot distinguish them. We believe that shows the excellent performance of our models.

\noindent \textbf{Space and Time Trade-off with a More Efficient 3D Implementation.}~In our current implementation of the 3D model, our focus is to demonstrate the effectiveness of our idea in both interpolation and extrapolation for 2D/3D scenarios. The performance for 3D scenarios is indeed worse than 2D because of the limitations of the computational resources in both memory and efficiency. When extending the model from 2D model to 3D, the major technical barriers would be the high computational training cost and the large memory consumption. Although the additional computational improvement for 3D scenarios is not the focus of this manuscript, we have already identified numerical solutions to overcome the limitations of computational resources. Particularly, we will leverage group convolution techniques to reduce the global convolution operations on the channel dimension, which has been proved to be effective in reducing the overall cost~\cite{howard2017mobilenets}. To reduce the memory consumption, we will trade storage for computation. It is well known that most of the memory consumption of a neural network models result from a large number of intermediate activations. Instead of saving all the activations, we will only store some activations and recompute those in-between when needed. This is described as in the ``Invertible  Network''~\cite{dinh2014nice}. 

\noindent \textbf{Known Knowns versus Unknown Unknowns.}~Extrapolation, in general, is challenging for any machine learning model including ours. However, what makes our problem unique and different comparing to many other problems is the existence of the underlying physics knowledge, which yields consistent spatio-temporal dynamics such as ``velocity push-down''. Our neural network models have been proved to be effective in capturing those characteristics. That in turn will help in the prediction of the near-future unknown data due to the invariant nature of the underlying physics. This is what we refer to as ``known knowns''. There will also be ``unknown unknowns'', which refers to uncertainty in the data. Provided with the limited availability of the existing data, the limitation of our model, and some additional unpredictable variances of the problem itself, the predictability of the machine learning models would become unattainable. That has also been reflected in our numerical testing results~(i.e., the increase of the loss values furthering out in the future). To improve the prediction performance, one immediate solution would be to fine-tune the model with newly acquired data.

\noindent \textbf{Generative Models and Others.}~Deep generative models~(DGMs), such as generative adversarial network~(GAN)~\cite{radford2015unsupervised, berthelot2018understanding}, variational autoencoder~(VAE)~\cite{kingma2014auto} and adversarial autoencoder~(AAE)~\cite{makhzani2015adversarial}, learn lower-dimensional data distribution explicitly or implicitly. Once the data distribution is successfully captured, DGMs can be used to generate synthetic manipulated data for interpolation. However, training DGMs requires a great amount of data, which would make them inappropriate for the interpolation and extrapolation of Sleipner data. For cases when a lot more 4D seismic datasets with much more time samples become available, DGMs would be interesting approaches to study. We plan to study their capability in spatio-temporal in-between image generation in the future. In our method, we utilize optical flow to track the temporal change of the seismic data over time and incorporate it to regularize the loss function. We will compare the performance using other motion tracking techniques, such as template matching~\cite{brunelli2009template}, dynamic warping~\cite{hale2013dynamic,feng2019transmission+} and kernelized correlation filter~\cite{henriques2014high}. In addition, it will be interesting to incorporate non-seismic data in the Sleipner area, such as gravity data and well-logging, into our proposed network to further regularize our model and potentially improve the results. The selection of the hyper-parameters (such as $\alpha^\mathrm{g}_{n}$, $\alpha^\mathrm{g}_{n+i}$, $\alpha^\mathrm{g}_{n+m}$ used in Eq.~\eqref{eq:loss} and $\alpha^\mathrm{e}_n$, $\alpha^\mathrm{e}_{n+i}$, $\alpha^\mathrm{e}_{n+2i}$, $\alpha^\mathrm{e}_{n+3i}$, $\alpha^\mathrm{e}_{n+4i}$ used in Eq.~\eqref{eq:loss_extra}) play an important role in producing accurate estimation. In this manuscript, we set the parameter based on their importance to the resulting prediction. Considering the relatively small size of our dataset, our empirical approach to select those hyper-parameters has been demonstrated to be rather effective via the numerical results. For a large dataset with more varying spatio-temporal dynamics, a more heuristic optimization technique would be needed to derive the optimal parameters. We would consider that as a future direction.

\section{Conclusions}
In this paper, we develop two neural-network-based methods for spatio-temporal in-between image interpolation and extrapolation. To capture the spatial representation, our models leverage the deep autoencoder to obtain semantically meaningful latent variables, which can be combined linearly to generate new features for interpolation/extrapolation tasks. Temporal dynamics are also important in designing our models. To enhance the temporal correlation of data, we utilize both the LSTM structure and optical flow regularization in our neural network models. We validate the performance of our models using 2D/3D post-stack seismic imaging data acquired at the Sleipner CO$_2$ sequestration field. Through both numerical comparisons to baseline methods and experts evaluation, we show our models yield high fidelity seismic imaging data that would allow in-situ reservoir monitoring and forecasting, and therefore yielding the great potential for CO$_2$ sequestration and storage.

\section{Acknowledgements}

This work was co-funded by the U.S. DOE Office of Fossil Energy's Carbon Storage program and by the Laboratory Directed Research and Development program of LANL under project numbers 20210542MFR and 20200061DR.

\bibliographystyle{IEEEtran}
\bibliography{main}

\end{document}